\documentclass[aps,pre,twocolumn,showpacs,superscriptaddress,groupedaddress,floatfix]{revtex4-2}
\usepackage{amsmath}
\usepackage{amssymb}
\usepackage{amsfonts}
\usepackage{graphicx}
\usepackage{dcolumn}
\pdfoutput= 1
\usepackage{sidecap}
\usepackage{parskip}
\usepackage{grffile}
\usepackage{color}
\usepackage{hyperref}
\usepackage{hhline}
\usepackage{mathtools}
\usepackage{graphics}
\usepackage{multirow}
\usepackage{verbatim}
\usepackage{longtable}
\usepackage{rotating}
\usepackage{setspace}
\usepackage{epsfig}
\usepackage{subfigure}
\usepackage{epstopdf}
\usepackage[normalem]{ulem}
\makeatletter
\def\@eqnnum{{\normalsize \normalcolor (\theequation)}}
 \makeatother

\graphicspath{{./}{ER/main/}}

\begin{document}

\title{Multiple first-order 
transitions in simplicial complexes on multilayer networks}

\author{Sarika Jalan and Ayushi Suman}
\affiliation{Complex Systems Lab, Department of Physics, Indian Institute of Technology Indore, Khandwa Road, Simrol, Indore-453552, India}

\begin{abstract}
The presence of higher-order interactions (simplicial complexes) in networks and certain types of multilayer networks has shown to lead to the abrupt first-order transition to synchronization. We discover that simplicial complexes on multilayer networks can yield multiple basins of attraction, leading to multiple routes to the abrupt first-order transition to synchronization.
Using the Ott-Antonsen approach, we develop an analytical framework for simplicial complexes on multilayer networks, which thoroughly explains the origin and stability of all possible dynamical states, including multiple synchronization transitions, of the associated coupled dynamics. 
The study illustrating rich dynamical behaviours could be pivotal to comprehending the impacts of higher-order interactions on dynamics of complex real-world networks, such as brain, social and technological, which have inherent multilayer networks architecture.

\end{abstract}

\maketitle

\section{Introduction}
Functional performances of a wide range of real-world complex systems, such as Brain, social and technological systems, are driven by underlying simplicial complexes defining higher-order interactions  \cite{rev_simplicial,simplicial_Ginestra2020,D_ghosh2022,HR_Perc}. A simplicial complex of order one denotes pair-wise interactions between a pair of nodes. Similarly, a simplicial of order two represents a set of three connected nodes forming a 2-simplicial complex, and so on. Considering such higher-order interactions,  aka simplicial complexes in interacting nonlinear dynamical units, has brought forward many emerging phenomena  \cite{rev_simplicial2021}. Among which, simplicial complexes have been elucidated to lead the abrupt first-order transition to synchronization is of particular interest. Precisely, the simplicial complexes of order two or higher have demonstrated to cause the first-order abrupt transition to synchronization in those systems which otherwise (i.e., only with pair-wise interactions) depict a smooth second-order transition to synchronization \cite{Skardal2019,Xu2020,Skardal2020,Ajaydeep2022}. 

Furthermore, a set of nodes of a complex system connected through different types of interactions form a multilayer network. Investigations of real-world data, as well as both the numerical simulations and theoretical analysis of nonlinear models, have demonstrated that in complex systems represented by multilayer networks, activities of nodes in one layer, connected through one type of interactions, affect or govern dynamical activities of the nodes connected through other types of interactions in the other layers \cite{Boccaletti2014,Pinaki2018,Anil_Inma2019,Frolov2021}. Similar to the first-order synchronization transition behaviour exhibited by coupled dynamics on  simplicial complexes,  appropriate multilayer schemes have also been shown to lead to the abrupt first-order transition to synchronization in those networks which, in isolation, depict smooth second-order transition to synchronization \cite{Anil_Inma2019,Anwar_Ghosh_2022,Ghosh_chimera_2022}. 
\begin{figure}[ht]
\includegraphics[width=0.48\textwidth]{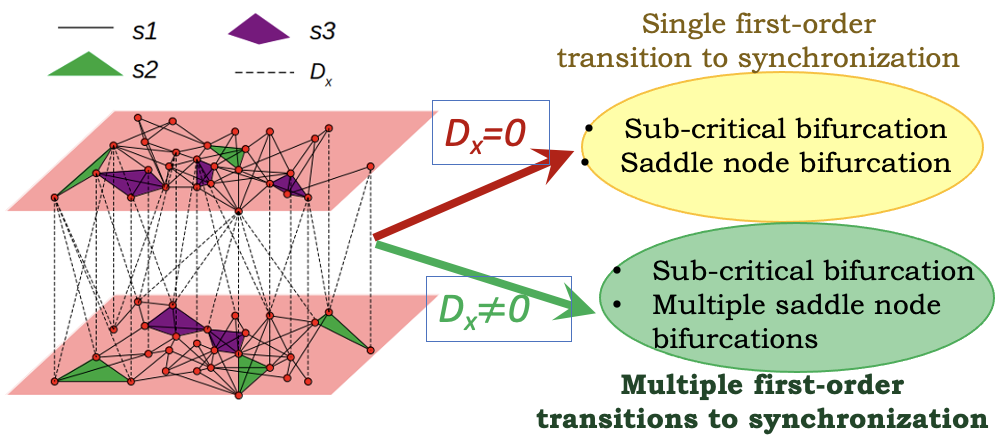}
\caption{(Colour online) Schematic diagram for (left) multilayer network consisting of simplicial layers, and (right) emerging dynamical phenomena in absence ($D_x=0$), and in presence  of multi-layering (($D_x>e0$).}
\label{fig:my_label} 
\end{figure}

This Letter discovers the existence of multiple basins of attractions, yielding different routes to abrupt first-order transitions to synchronization in simplicial complexes on multilayer networks. Moreover, we develop a complete analytical framework using the Ott-Antonsen approach, where we analytically deduce the high dimensional multilayer coupled dynamics to a low dimensional coupled equations for the associated order parameters of the individual simplicial layers. Such reduced time-dependent coupled equations for the simplicial layer's order parameters facilitate
analytical demonstration of different routes to first-order transitions and calculations of different critical coupling strengths at which these transitions occur.   
Notably, using the Ott-Antonsen approach, we first derive the time evolution equations for the order parameters of the simplicial complexes of coupled Kuramoto oscillators having pair-wise ($s1$), triadic-($s2$) and tetra-($s3$) interactions forming layers of the multilayer networks. After that, we perform stability analysis for the set of the coupled differential equations for the order parameters of the layers at all their fixed points. The analytical calculations reveal the existence of a usual one sub-critical bifurcation leading to the first-order transition to synchronization, which is accompanied by hysteresis, and an abrupt transition to synchronization as a consequence of saddle-node bifurcation. Notably, depending upon a choice of the initial conditions, the set-up leads to three additional saddle-node bifurcations, giving birth to another second-order transition to synchronization accompanied by a different critical coupling strength. We perform numerical experiments which comply with the analytical predictions. 


\section{Model and Results}
Let us consider multilayer networks consisting of two layers without loss of generality. The $N$ number of nodes in each layer interacts not only with the pair-wise interactions (simplicial of order 1) but also through triangular (2-simplicial) and tetrahedral (3-simplicial) interactions (Fig.~1). The inter-layer pair-wise interactions are considered all-to-all, i.e., each node of the first simplicial layer interacts with all the nodes of the second simplicial layer with a strength $D_x$, referred to as multilayering strength. The dynamics of such a system can be studied with an extension of the Kuramoto-Sakaguchi model
\cite{Kura1984} given as
\begin{equation}
	\dot{\theta_{i}}^{(k)} =\omega_{i}^{(k)}+\frac{D_{x}}{N}\sum\limits_{j=1}{\sin(\theta_{j}^{(k')}-\theta_{i}^{(k)})}+\hat{H}_{i}^{(k)}
	\label{eq_coup}
	\end{equation}
where, 
\begin{eqnarray}
\hat{H}_i^{(k)} &=&\frac{K_1^{(k)}}{N}\sum\limits_{j=1}\sin(\theta_j^{(k)}-\theta_i^{(k)}) \nonumber\\
&+&\frac{K_2^{(k)}}{N^2} \sum\limits_{j=1}\sum\limits_{l=1}\sin(2\theta_j^{(k)}-\theta_l^{(k)}-\theta_i^{(k)})  \nonumber\\
&+&\frac{K_3^{(k)}}{N^3} \sum\limits_{j=1}\sum\limits_{l=1} \sum\limits_{m=1}\sin(\theta_j^{(k)}+\theta_l^{(k)}-\theta_m^{(k)}-\theta_i^{(i)})\nonumber
\end{eqnarray}
Here, the superscript $k$, taking values 1 and 2,  denotes the index of the layer in the consideration, and $K_i^{(k)}$ indicates  the  overall coupling strength for the $i$-simplex  interaction of the $k^{th}$ layer.  In the respective mean fields, the dynamical evolution equations for the simplicial layers can be written as
\begin{eqnarray}
	\dot{\theta_i}^{(k)} &=&\omega_i^{(k)}+K_1^{(k)}r_1^{(k)}\sin(\Psi_1^{(k)}-\theta_i^{(k)}) \nonumber\\ &+&K_2^{(k)}r_1^{(k)}r_2^{(k)}\sin(\Psi_2^{(k)}-\Psi_1^{(k)}-\theta_i^{(k)}) \nonumber \\ &+&K_3^{(k)}r_1^{(k)}\sin(\Psi_1^{(k)}-\theta_i^{(k)}) \nonumber \\ &+& D_x \sum_{m\ne k}r_1^{(m)} \sin(\psi_1^{(m)}-\theta_i^{(k)}),
	\label{eq_mean}
	\end{eqnarray}
with the complex order parameters of the $k^{th}$ simplicial layer defined as, 
\begin{equation}
 z^{(k)}= 
r^{(k)}e^{i\Psi^{(k)}}=\frac{1}{N}\sum\limits_{j=1}e^{i\theta_j^{(k)}}
\label{eq_order}
\end{equation}
which measures the strength of global synchronization of the individual simplicial layers. $r_1^{(k)} \sim 0$ indicates a complete incoherent state, whereas $r_1^{(k)} \sim 1$ indicates global synchronization of the oscillators of the $k^{th}$ layer. In the absence of any higher-order interaction, Eq.~1 leads to a smooth second-order transition to synchronization for the individual layer \cite{Anil_Inma2019}, i.e. starting from an incoherent state corresponding to  $r^{(k)} \sim 0$, after a critical coupling strength $r^k$ gradually increases to 1 with an increase in $K_1$. An introduction of the higher-order couplings (second and third term in $\hat{H}_i$) brings about frustration in the system due to an interplay among the phases of three and four oscillators, respectively, for the triadic and tetrahedral term, in the sinusoidal couplings. As the pair-wise intra-layer coupling strength increases, it supports the overall coherence between the connected pairs of nodes. As soon as $K_1$ becomes strong enough to dominate the frustration caused by the higher-order interactions, there arises an abrupt first-order transition to the global synchronization in the layer, witnessing the increase in the intra-layer pair-wise coupling strength. For the single-layer networks, the implications of the inclusion of higher-order interactions in coupled Kuramoto oscillators interacting through simplicial complexes on single-layer networks is a well-established model, accompanied by rigorous analytical explanation \cite{Skardal2020}. Here, we consider systems in which simplicial layers are connected via inter-layer connections and develop a full analytical framework to analyze the dynamical behaviour of coupled Kuramoto oscillators on simplicial complexes on multilayer networks. We discover various emerging behaviours, namely, multiple first-order transitions to synchronization accompanied by different routes and basins of attraction, the existence of multiple saddle-node bifurcations in addition to usual sub-critical pitch-fork bifurcation, and a gradual shift from the first- to the second-order transition to synchronization.

In the following, first, we develop a rigorous analytical framework for the complete accomplishment of the coupled dynamical behaviours and their stability analyses. Then, we present numerical results obtained through the direct simulations of Eq.~\ref{eq_coup} to demonstrate a full match with the analytical calculations. We perform a linear stability analysis of the reduced equations (\ref{eq_reduced}) and explore the entire $r-\rho$ space. Further, we discuss the origin of the third bifurcation, followed by an analysis of the impacts of intra-layer coupling strength of the second simplicial layer and inter-layer coupling strengths on the nature of transitions in the first layer. Next, we present results for the dynamical evolution of the second layer and discuss how the origin of multiple transitions lies in the dynamical behaviours of the layer having only the higher-order interactions.
\\

\paragraph{\bf{Analytical derivation:}} Using the complex order parameters (\ref{eq_order}), Eq.~\ref{eq_mean} can be written as 
\begin{equation}
 \dot{\theta_i}^{(k)}=\omega_i^{(k)}+\frac{1}{N}(H^{(k)}e^{-i\theta_i^{(k)}}-H^{*(k)}e^{i\theta_i^{(k)}})  
 \label{eq_mean2}
\end{equation}
 with $H^{(k)}=K_1^{(k)} z_1^{(k)}+K_2^{(k)} z_2^{(k)} z_1^{*(k)}+ K_3^{(k)} z_1^2 z^{*(k)}+ \sum_{m\ne k} D_x z_1^{(m)}$.
 In the thermodynamic limit $N \longrightarrow \infty$, the state of the system can be described by a density function $f_k(\theta^{(k)},\omega^{(k)},t)$ which measures the density of oscillators with phase between $\theta$ and $\theta + d\theta$ having natural frequency lying between $\omega$ and $\omega + d\omega$ at time $t$ for the $k^{th}$ layer. 
 Since the number of oscillators in each layer is conserved, the density functions will individually satisfy the continuity equation,
\begin{equation}
    0=\frac{\partial f_k}{\partial t}+\frac{\partial}{\partial \theta^{(k)}} \left[f_k \left[ \omega_i^{(k)} + \frac{1}{N} (He^{-i\theta_i^{(k)}}-H^*e^{i\theta_i^{(k)}}) \right] \right]
    \label{eq_continuity}
\end{equation}
Assuming the natural frequency $\omega$ of each oscillator drawn from a distribution $g(\omega)$, the density function $f_k$ can be expanded into Fourier series as
\begin{equation}
    f_k(\theta,\omega,t)=\frac{g(\omega)}{2\pi} \left[1+\sum\limits{n=1}\hat{f_n}(\omega,t)e^{in\theta} + c.c. \right],\nonumber
    \end{equation}
where $\hat{f_n}(\omega,t)$ is the $n^{th}$ Fourier component and c.c. are the complex conjugate of the former terms. Next, we use the Ott-Antonsen \cite{Ott_Antonsen} ansatz which assumes that all the Fourier modes decay geometrically, i.e., $\hat{f_n}(\omega,t)=\alpha^n(\omega,t)$ for some function $\alpha$ which is analytic in the complex $\omega$ plane. After inserting the Fourier expansion of $f_k(\theta,\omega,t)$ in the continuity equation (\ref{eq_continuity}), the dynamics of the two-layer network collapses into a complex two dimensional manifold (Ott-Antonsen manifold).
\begin{equation}
    \dot{\alpha}^{(k)}=-i\omega^{(k)}\alpha^{(k)}+\frac{1}{2} \left[H^{(k)*}-H^{(k)}\alpha^2  \right]
    \label{eq_alpha}
\end{equation} 
with $H^{(k)}$ defined in Eq.~\ref{eq_mean2}. 
The order parameter in the thermodynamic limit can then be given as $z^{(k)}$=$\int \int f^{(k)}(\theta^{(k)},\omega^{(k)},t)e^{i\theta^{(k)}}d\theta^{(k)} d\omega^{(k)}$, which after inserting the Fourier decomposition of $f^{(k)}$ becomes,
\begin{equation}
    z^{(k)}=\int \alpha^{(k)}(\omega^{(k)},t)g(\omega^{(k)})d\omega
    \nonumber
    \end{equation}
    If we choose $g(\omega)$ to be a Lorentzian frequency distribution $g(\omega)=\frac{\Delta}{\pi\left[\Delta^2+(\omega-\omega_0)^2\right]}$, where $\omega_0$ is mean and $2\Delta$ is full width at half maximum, $z^*$ can be calculated by contour integration in the negative half complex plane, yielding, $z^{*(k)}=\alpha^{(k)}(\omega_0-i\Delta,t)$.
With $z^{(1)}=re^{i\Phi}$ and $z^{(2)}=\rho e^{i\chi}$, 
and scaling $\chi-\Psi$ as $\xi$, dimensionality of the system represented by Eq.~\ref{eq_alpha} reduces to three,
{\small{
\begin{align}
	\begin{split}
	\dot{r} =-{\Delta}r+\frac{K^{(1)}_1}{2}\big[r-r^3]+\frac{K^{(1)}_{2+3}}{2}\big[r^3-r^5]+\frac{D}{2}\big[\rho-\rho r^2]\cos(\xi),\nonumber
	\end{split}\\
	\begin{split}
	\dot{\rho} = -{\Delta}\rho+\frac{K^{(2)}_1}{2}\big[\rho-\rho^3]+\frac{K^{(2)}_{2+3}}{2}\big[\rho^3-\rho^5]+\frac{D}{2}\big[r-r\rho^2]\cos(\xi),\nonumber
	\end{split}\\
	\begin{split}
	    \dot{\xi}=\omega_0^{(2)}-\omega_0^{(1)}-\frac{D}{2}\bigg[2r\rho+\frac{r}{\rho}+\frac{\rho}{r}\bigg]\sin(\xi)
	\end{split}
	\label{eq_reduced1}
\end{align}}}
In  the steady state, $\dot{r}=\dot{\rho}=\dot{\xi}=0$. Note that since this analytical derivation  has considered   the Cauchy frequency distribution centered at zero,  for $\dot{\xi}$ to be zero, $sin(\xi)$ has to be zero (since the quantity in square bracket cannot be zero). Which means $\chi=\Psi$, indicating synchronization of the mean phases of both the layers. Consequently, the above three coupled equations further reduce to 
\begin{align}
	\begin{split}
	0 =-{\Delta}r+\frac{K^{(1)}_1}{2}\big[r-r^3]+\frac{K^{(1)}_{2+3}}{2}\big[r^3-r^5]+\frac{D}{2}\big[\rho-\rho r^2],\nonumber
	\end{split}\\
	\begin{split}
	0 = -{\Delta}\rho+\frac{K^{(2)}_1}{2}\big[\rho-\rho^3]+\frac{K^{(2)}_{2+3}}{2}\big[\rho^3-\rho^5]+\frac{D}{2}\big[r-r\rho^2]	\label{eq_reduced}
	\end{split}
\end{align}
Eq.~\ref{eq_reduced} provides solution to the dynamically  stable states for the multilayer networks model (\ref{eq_coup}) in terms of the order parameters of the individual simplicial layers ($r$ and $\rho$). Ergo, a high dimensional coupled dynamics is reduced to 2-dimensional evolution equation.
This set of the coupled equations provides full understanding to the entire dynamics of the multi-layer network model represented by Eq.~\ref{eq_coup}. 

\begin{figure}
\includegraphics[width=0.5\textwidth]{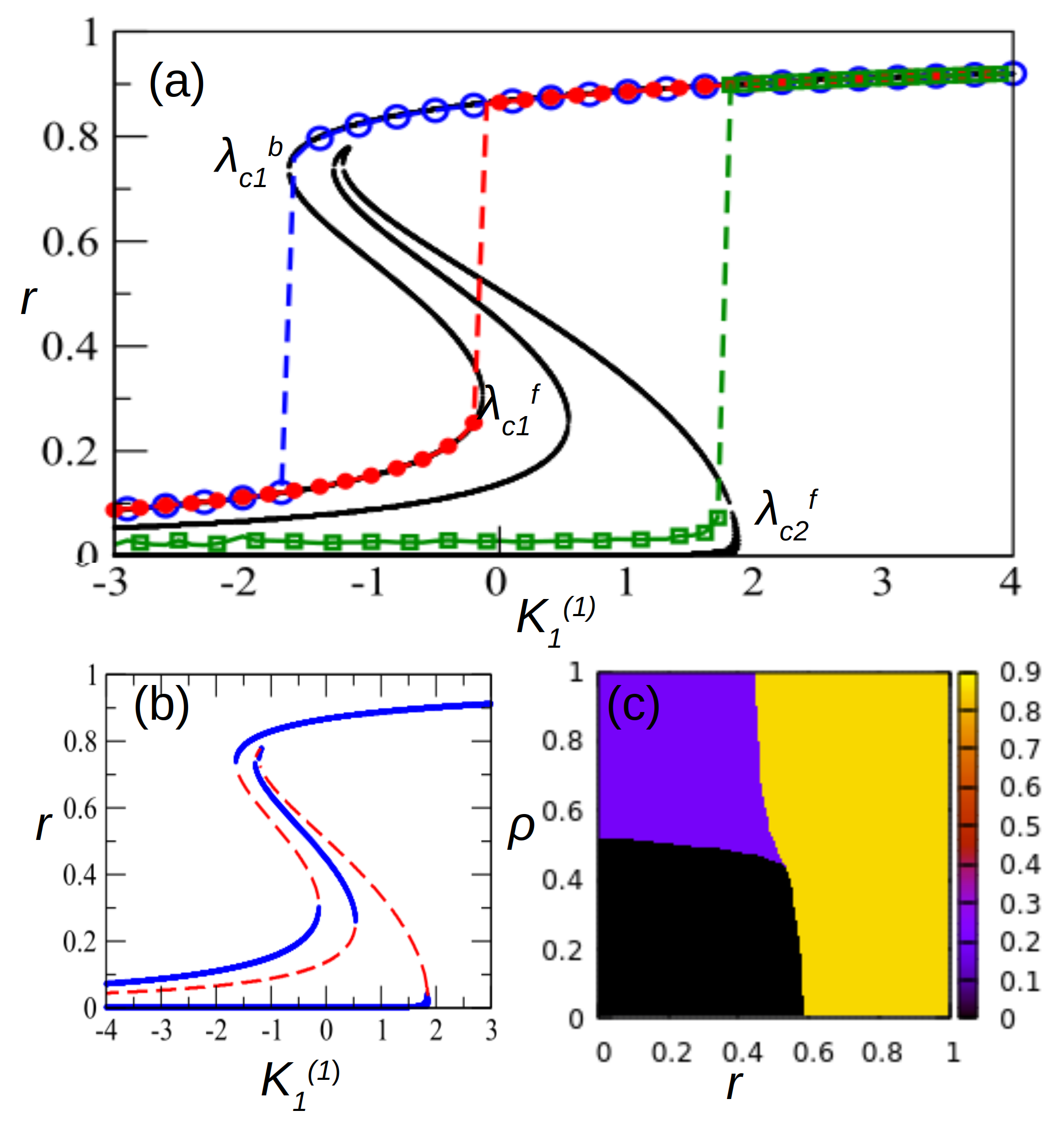}
\caption{(Colour online) Multiple synchronization transitions or multi-stability along with analytically predicted curves. (a) Open  circles (blue) indicate adiabatic backward while closed circles (red) represent adiabatic forward direction, both combined making HT (see text). Open squares (green) indicate forward direction corresponding to the random initial conditions, also referred as WHT (see text). Solid lines are analytical solutions of Eq.~\ref{eq_reduced}.  (b) stability diagram: Solid (blue) and dashed (red) lines, respectively, indicate stable and unstable manifolds. (c) Basins of attraction for the entire  $r-\rho$ space depicting three stable regions for Eq.~\ref{eq_reduced1} for all  possible initial  conditions for $r$ and $\rho$. At $K_1^{(1)}=-1$, black, grey (purple) and white (yellow) correspond to the green open square, red closed circle and blue open circle, respectively, in (a).}

\label{fig:my_label} 
\end{figure}

\paragraph{\bf{Multiple first-order transitions and basins of attraction:}}
The coupled dynamics (Eq.~\ref{eq_coup}) is evolved using the Runge-Kutta method of order 4, and after discarding an initial transient, the system settles to a stable state. That stable state is investigated using the order parameter defined by Eq.~\ref{eq_mean}.
Additionally,  Eq.~7 is solved numerically, and steady-state behaviour of $r$ and $\rho$ are analyzed as a function of $K_1^{(1)}$. To comprehend the dynamical behaviour of the nodes in one simplicial layer on those of the other layer, we freeze the coupling strengths of one layer (say, the second layer, $K^{(2)}_1$ and $K^{(2)}_{2+3}$) to a set of values, and study the dynamical evolution of the nodes of another layer (say the first layer). The solid black lines in Fig.~2(a) correspond to the theoretical predictions based on the Ott-Antonsen method (\ref{eq_reduced}), which matches very well with the numerical evolution of the system (\ref{eq_coup}). For the multilayer simplicial networks, We discover the existence of multiple stable states accompanied by different basins of attraction which yield different routes to the first-order transition to synchronization.
First, we discuss the 
 numerical results for the two-layer networks having all-to-all intra-layer couplings.
Different routes to the first-order transitions and various bifurcations exist depending upon the initial conditions.
(1) 
Phase of the oscillators in both the layers are randomly distributed in $[0:\pi]$. For this set of initial conditions and  $K_1^{(1)}$, $K^{(1)}_2$, $K^{(1)}_3$ taking any positive small values leads to $r=0$ as a stable solution. Upon increasing $K^{(1)}_1$, there exists a usual first-order  transition to synchronization at a critical coupling strength ($K_1^{(1)} = \lambda_{c2}^f$)) as a consequence of the sub-critical bifurcation as also observed for isolated simplicial layer case \cite{Skardal2020}. 
(2) Initial condition corresponding to the  synchronized state ($\theta_{i} = \theta_{j}, \forall i,$) which  is achieved for sufficiently large $K_{1}^{(1)}$ values. Starting with this initial condition, as $K_{1}^{(1)}$ decreases adiabatically,  the dynamical evolution of the first simplicial layer lies in the coherent region (blue open circles) until the critical coupling strength $(K_{1}^{(1)}=$ $\lambda_{c1}^b$), where the stable, coherent state losses its stability as a consequence of a saddle-node bifurcation. The trajectory jumps to a far distant attractor corresponding to an incoherent state (blue open circles). The incoherent state persists as $K_{1}^{(1)}$ decreases further. 
(3) Initial condition corresponding to the incoherent state (blue open circles), which was achieved in the backward direction:  Upon increasing $K^{(1)}_1$ adiabatically from this initial condition, the order parameter trajectory takes a different route than was followed while $K^{(1)}_1$ was decreasing. The coupled dynamics keeps on depicting the incoherent behaviour beyond $\lambda_{c1}^b$ until specific values of $K^{(1)}_1$ (red closed circles). At $\lambda_{c1}^f$, this incoherent stable state losses its stability
again as a consequence of a saddle-node bifurcation and giving birth to the first-order transition to the coherent state. 

Since this study demonstrates the existence of multiple synchronization transitions as the prime result, we refer to the transition corresponding to the hysteresis as HT and the transition without hysteresis as WHT. Therefore, in Fig.~2(a), blue open circles and solid lines correspond to backward HT, red closed circles and dashed lines correspond to forward HT, whereas green open squares correspond to WHT.

\paragraph{\bf {Origin of the third bifurcation:}}
Eq.~\ref{eq_reduced} is numerically solved to obtain all the bifurcating lines depicted in Fig.~2(b). Furthermore, we perform the linear stability analysis around all the fixed points of Eq.~6 by evaluating the respective Jacobian matrices. A fixed point is asymptotically stable if both the eigenvalues of the corresponding Jacobian matrix have negative real parts (solid blue line in Fig.~2(b)), and it is unstable if at least one eigenvalue has a positive real part (red dashed line in Fig.~2(b).
In addition to the stable manifold observed numerically (blue solid line), there exist three unstable manifolds (red dashed line), and one more stable manifold (blue solid line in the middle of two dashed red lines) which were not observed numerically. 
Upon analyzing a cross-section at $\xi=0$  corresponding to $K^{(1)}_1=-0.5$, one finds the existence of four stable manifolds. However, one stable state here corresponds to the out of the phase-synchronized solution between the layers and is unstable in the $\xi$ direction. Therefore, this curve is not realized numerically from the reduced coupled mean-field equations, which is solved for the $\xi=0$. 
Fig.~2(c) plots the basins of attraction for different regions, in which only three stable regions are visible for $K^{(1)}_1=-1$. (c) is plotted in the $r$-$\rho$ parameter regime. To obtain the initial conditions for phases corresponding to a specific set of values of $r$ or $\rho$, one can use an asymmetry parameter $\eta$ such that $N\eta$ oscillators start at the initial phase $0$ and $N(1-\eta)$ oscillators start at the initial phase $\pi$. Then $\eta=(r,\rho+1)/2$.

\begin{figure}
\includegraphics[width=0.5\textwidth]{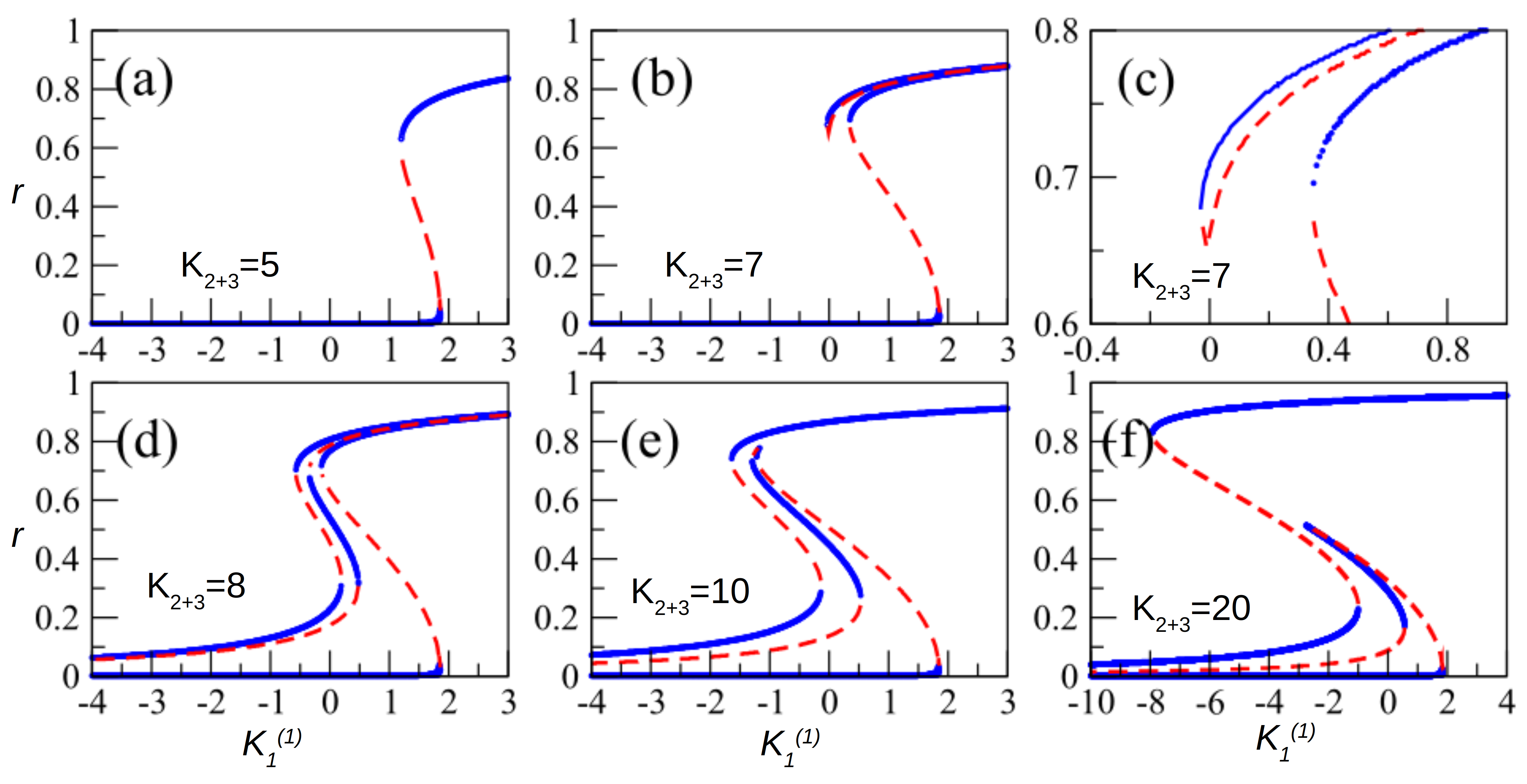}
\caption{(Colour online) Effect of higher-order interactions strength:  $K_{1}^{(1)}$ vs $r$.
Sub-figures illustrate the emergence and disappearance of a saddle node bifurcation leading to the emergence of the third stable state in the system for different values of $K_{2+3}$. 
(c) Zoomed part of (b) corresponding to the regime where a new branch emerges. Here $D_{x}=1$, and $K_{1}^{(2)}=0$.}
\label{fig:my_label}
\end{figure}

\begin{figure}[h]
\includegraphics[width=0.5\textwidth]{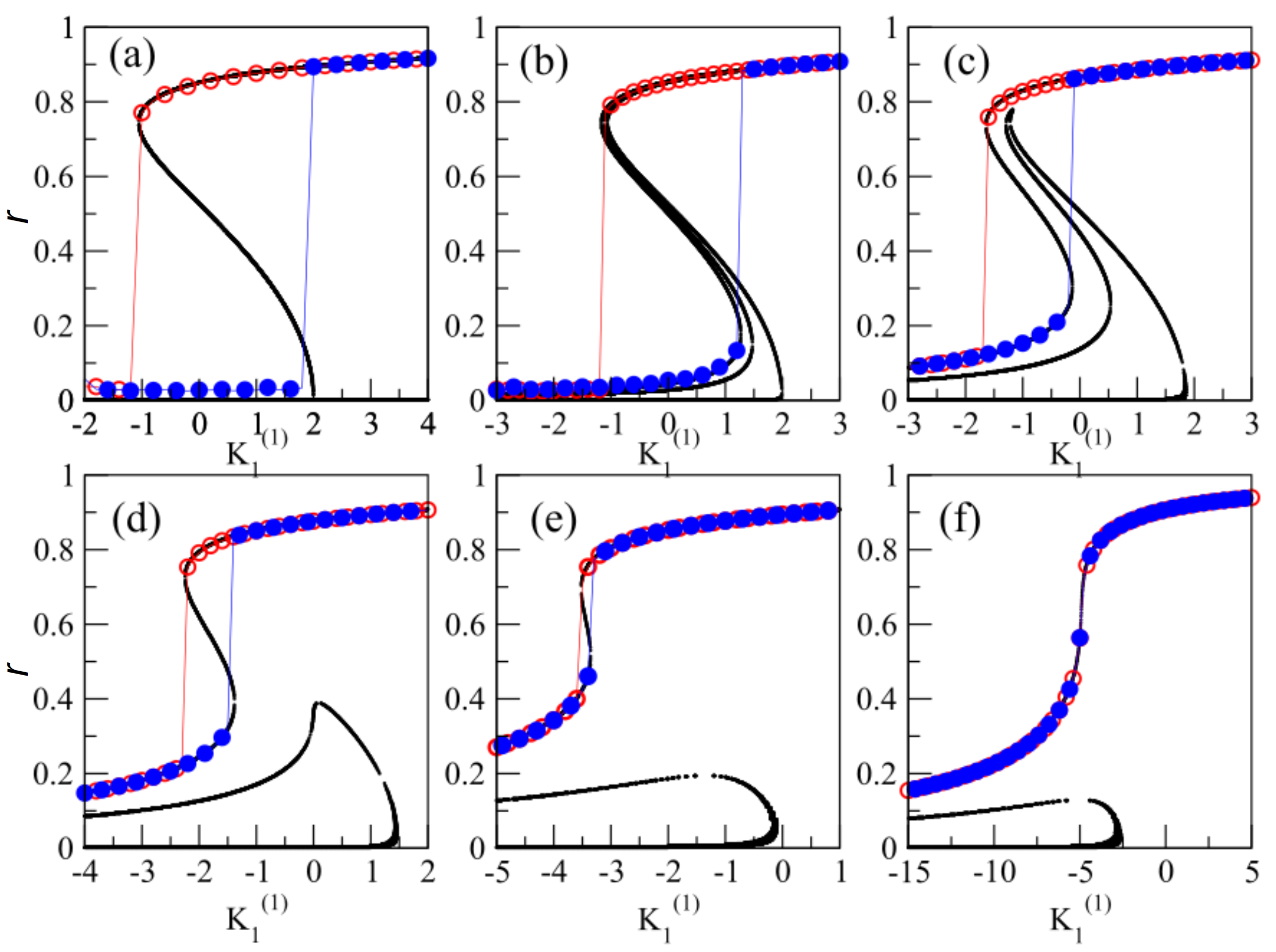}
\caption{(Colour online) Loss of hysteresis as a result of an increase in $D_{x}$.  $r^{1}$ vs. $K_{1}^{(1)}$ for $K_{2+3}=10$. (a)-(f) Backward (red open circle) and forward (blue closed circle) transitions for $D_{x}$ values 0, 0.1, 0.5, 1, 2 and 3, respectively. The black solid curve represents theoretical prediction using Ott-Antonsen method (\ref{eq_reduced}). }
\end{figure}

\paragraph{\bf{Impact of 2- and 3- simplex interaction strengths:}}
We next explore the impact of change in the strengths of higher-order interactions on the nature of the dynamical evaluations, transitions points and different basins of attraction. Fig.~3 illustrates the emergence and disappearance of the stable branches in $K_{1}^{(1)}-r$ space for different values of $K^{(1)}_{2+3}=K^{(2)}_{2+3}=K_{2+3}$. For smaller values of $K_{2+3}$, there exists only one usual first-order transition route to synchronization as a function of pair-wise couplings (blue solid line in (a)). This transition arises due to a sub-critical pitch-fork bifurcation leading to the bi-stable state for the simplicial layer. As $K_{2+3}$  increases, a new pair of stable-unstable states emerges (Figs.~3(b) and (c)), which runs in very close vicinity (almost indistinguishable) to the existing stable coherent branch.
Upon increasing $K_{2+3}$ and further for both the simplicial layers simultaneously, this new emerged branch which was running along the existing stable curve, joins the incoherent state gradually for large values of $K_{2+3}$ (Fig.~3(d)). Indeed, the previously existing stable branch gradually gets separated from the new emerging one (Figs.~3(e)-(f)), and then vanishes (not shown here), again restoring the same stability diagram with only one first-order transition (hysteresis) as found for the lower $K_{2+3}$ values. 
Note that the figure is plotted only  for a particular parameter values ($D_{x}=1$ and $K_{1}^{(2)}=0$). However, this existence of multiple synchronization transitions is witnessed for other lower values of $K_{1}^{(2)}$, and also as long as  a threshold value of $K_{2+3}$ is crossed. 
\\

\paragraph{\bf{Impact of $D_x$: First- to second-order transition}}
Next, we analyze the effect of   inter-layer coupling strength $D_{x}$  on the dynamical evolution of the individual simplicial layers. For lower values of $D_{x}$, two stable states near $r=0$ lie in very close vicinity to each other and are indistinguishable (Figs.~4 (a)-(b)). The basin of attraction for the stable incoherent state corresponding to $r=0$ is relatively small and difficult to achieve numerically. 
As $D_{x}$ increases gradually, two events take place; first, the hysteresis width becomes smaller (Figs.~4(a)-(e)) and eventually vanishes, leading to a smooth second-order transition from the incoherent to the coherent state (Fig.~4(f)). The larger $D_{x}$ values intensify the impact of pair-wise interactions in the entire two-layer networks, driving the second-order transition to synchronization for the simplicial layers. Since the pair-wise couplings are known to favour synchronization among the connected nodes, or in other words, for pure $s1$ complex,  there exists an emergence of a giant cluster with an increase in the pair-wise coupling strength. This cluster attracts more and more nodes leading to a second-order transition to synchronization \cite{Strogatz_Kura}. For all the nodes in the first layer connected to all the nodes in the second layer, an increase in $D_{x}$ readily dominates the interplay between the pair-wise interactions and the 2-, 3- simplex interactions. As a result, for large $D_{x}$ values, the transition becomes purely of the second-order. Second, all the transition points move towards more negative $K^{(1)}_1$ values as $D_{x}$ increases, i.e., synchronization occurs for a more extensive range of $K^{(1)}_1$. 
\begin{figure}[h]
\includegraphics[width=0.5\textwidth]{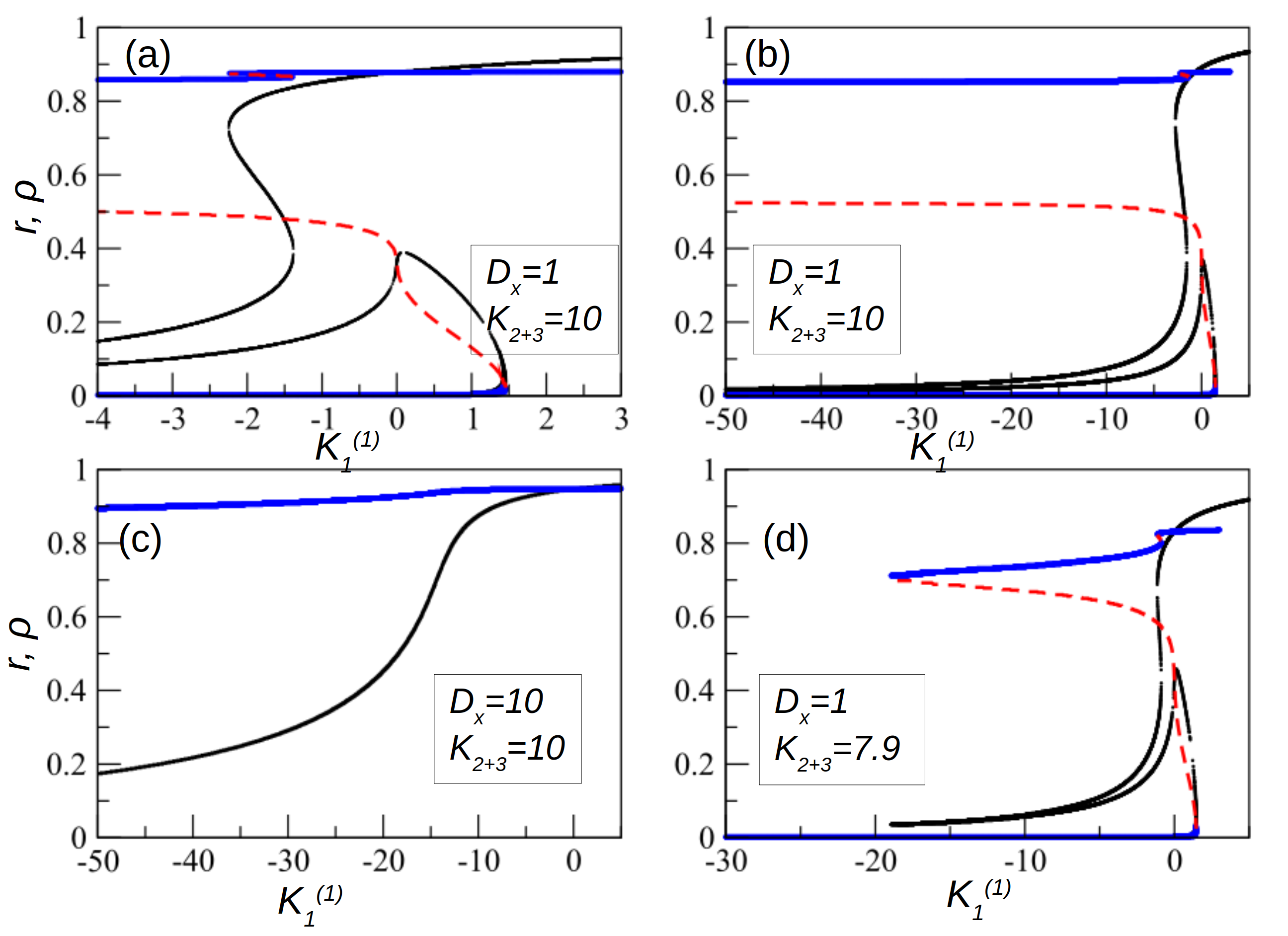}
\caption{(Colour online)(a) Synchronization profile for the second layer (red dashed line  for stable and thick blue solid for unstable manifold) as compared to the first layer (black thin line) for both stable and unstable manifolds for different values of $D_x$ and $K^{(2)}_{2+3}$. Robustness of behaviour of $\rho$ against  change in (b) $K_1^{(1)}$, and (c) $D_x$.  (d) Dependence of $\rho$ on $K_{2+3}^{(2)}$. For $K_{2+3}^{(2)}=7.9$, the pure simplicial layer de-synchronizes at rather a very high negative value of $K_1^{(1)}$. The range of $K_1^{(1)}$ for which the pure simplicial layer does not  get de-synchronized increases with further increase in  $K_{2+3}^{(2)}$ values (a)-(c).}
\label{fig:fig_second_layer}
\end{figure} 
\\

\paragraph{\bf {Dynamical evolution of the second layer:}}
Here, we systematically analyze the importance of higher-order interactions ($s2$ and $s3$)  in one layer in deciding a particular dynamical behaviour of another layer. We demonstrate that the second layer plays a governing role in deciding the nature of the synchronization transition in the first layer. For the second layer consisting of $s2$ and $s3$ only, the setup leads to one more stable state in the first layer, the origin of which lies in the dynamical behaviour of the second simplicial layer. Starting from  $K^{(1)}_1$ values for which both the simplicial layers are in the synchronized state, as $K^{(1)}_1$ decreases, while the first layer experiences a  transition to an incoherent state,  the second layer, despite being extensively connected to the first layer through the pair-wise inter-layer connections, does not get de-synchronized (Fig.~5(a)). Next, starting with the initial condition for the phases randomly distributed between $0$ and $\pi$, as $K^{(1)}_1$ increases, blue thick and black thin lines overlap with each other and together, they exhibit a jump to a coherent state at $K^{(1)}_1~1.5$ (black lines corresponding to the first layer, explained in the previous section). Next, Fig.~5(b) extends $K^{(1)}_1$ axis incorporating more negative values to demonstrate the robustness of the coherent behaviour of the second layer against change in $K_{1}^{(1)}$ values. Also, the dynamical behaviour of the second simplicial layer is robust against the changes in the $D_{x}$. For $D_{x}$ being as high as 10, the simplicial layer does not experience any de-synchronization transition even for very high negative $K^{(1)}$ values (Fig.~5(c)). Only for very small $K_{2+3}$ values the pure simplicial layer gets de-synchronized soon after the pair-wise interactions in the first layer cross the origin, and that too, this de-synchronization happens much after the first layer has de-synchronized (Fig.~5(d)). 
Therefore, it can be concluded that dynamical behaviours of a layer consisting of pure higher-order interactions are robust against changes in the multilayering parameters.
\label{fig:my_label}

\paragraph{\bf {Role of pure simplicial layer in the birth of multiple transitions:}}
The absence of pair-wise interactions, referred to as a pure simplicial layer, causes the multiple transition routes for the order parameter of the first layer. Starting from the coherent state (green open square, Fig.~2(a)), upon adiabatically decreasing $K^{(1)}_1$, as the first layer undergoes a first-order transition to the incoherent state (blue open circle, Fig.~2(a)), the second layer also experiences a slight jump to a state that still is a coherent state (Fig.~5(a)). Thereafter, the evolution of the Eq.~1 with a set of the initial conditions for which the first layer lies in the incoherent regime and the second layer in the coherent state, with an adiabatic increase in $K_1^{(1)}$, shows hysteresis and bi-stability (red closed circles of Fig.~2(a)) as a  consequence of the subcritical pitchfork bifurcation at $\lambda^f_{c2}$. At this critical point, the first layer jumps to the coherent state, and the second layer also keeps lying in the coherent state; ergo, both the layers evolve synchronously.

\section{Conclusion}
In summary, we have reported, analytically and numerically, the emergence of new stable manifolds as a consequence of connecting a simplicial layer to another simplicial layer. Analytically, by employing the Ott-Antonsen approach, we reduce the dimensionality of the two-layer networks to a set of two coupled differential equations of the order parameters of the individual layer. Such a reduction facilitates a straightforward way to comprehend the entire phase space fully and access all the bifurcations points and phase transitions. We found that an interplay of inter-layer strength and higher-order interactions brings forward several emerging dynamical behaviours, such as multiple routes to first-order transitions accompanied by corresponding basins of attraction. With the rigorous analytical calculations,  we demonstrate a rich phase space structure consisting of several stable and unstable manifolds arising from a sub-critical pitch-fork and three saddle-node bifurcations. Notably, after a critical higher-order interactions strength value,   connecting a simplicial layer with another simplicial layer via $D_x$ brings upon one more stable route to the first-order transition to synchronization, in addition to the previously existing route for the isolated simplicial layer.
Further, a simplicial network without pair-wise couplings does not depict a forward transition to synchronization if the initial conditions for phases are randomly drawn. Instead, it depicts a first-order transition to de-synchronization in the backward direction \cite{Skardal2019}. This article has revealed that connecting such a simplicial layer to another simplicial layer brings robustness to the synchronization in the backward direction. In the absence of pair-wise interactions ($K_1^{(2)}=0$), it does not get de-synchronized for the value of the parameter for which another simplicial layer, having pair-wise interactions, exhibits de-synchronization as a function of pair-wise coupling strength in the reverse direction.

The current work has only considered globally coupled networks for both the simplicial layers, whereas real-world systems have complex underlying network architectures. A more realistic model should include such a feature among many other real-world systems-inspired features, which is one of the future scopes of the present model. One could also relax one to all multilayer connections setup and consider a more general scheme represented by a multiplex matrix \cite{Ajaydeep_explosive_chimera2021,supra_lap2013}. Further, one can include various adaptive schemes \cite{Aoki2009,Berner2020,Berner2021} for the simplicial interactions, for example,  those inspired by the Hebbian learning in Brain \cite{SJ_Ajaydeep_NJP2022}.

These findings have applications in a range of systems having inherent multilayer architectures. It has been sufficiently emphasized that ignoring the impact of activities of nodes in one layer, having one type of interaction, may have wrong or inaccurate predictions of the dynamical evolution of the nodes interacting with another type of interaction. Using coupled Kuramoto oscillators on simplicial multilayer networks as a prototype model, based on rigorous analytical calculations supported by numerical simulations, we have provided evidence of emerging behaviours. So far, all the investigations and analytical calculations on multilayer networks have considered pair-wise interactions in their layers. This Letter develops a rigours analytical framework for simplicial complexes on multilayer networks. In addition, the numerical simulations reveal emerging dynamical behaviours, notably multiple first-order abrupt transitions to synchronization. A  range of real-world complex systems, such as  Brain, social \cite{Karsai2021, commphy2021}, financial and technological systems, have inherent multilayer network architecture \cite{rev_multi}, and their constituents units (nodes) interact with higher-order interactions forming simplex complexes  \cite{rev_higherorder}. The results presented in this article,  particularly the revelation of the occurrence of multiple transitions in simplicial complexes on multilayer networks, could be pivotal in predicting and comprehending the dynamics of such systems.

\begin{acknowledgements}
SJ gratefully acknowledges DST POWER grant SERB/F/9035/2021-2022. We thank Saptarshi Ghosh, Ajaydeep Kachhvah and complex systems lab members for useful suggestions.  \end{acknowledgements}


\begin{thebibliography}{99}


\bibitem{rev_simplicial}
    L. V. Gambuzza {\it et. al.}, {\em Nature Communications} {\bf 12}, 1255 (2021).
    
    \bibitem{simplicial_Ginestra2020}
A. P. Mill{\'a}n, J. J. Torres and G. Bianconi {\em Physical Review Letter} {\bf 124}, 218301 (2020)..
    
\bibitem{D_ghosh2022} S. Majhi, M. Perc, and D. Ghosh,  {\em J. R. Soc. Interface} {\bf 19}, 0043 (2022).

\bibitem{HR_Perc}F. Parastesh, {\it et. al.},
{\em Chaos} {\bf 32}, 013125 (2022). 

\bibitem{rev_simplicial2021} F. Battiston {\it et. al.}, {\em Nat. Phys.} {\bf 17}, 10 (2021).

\bibitem{Skardal2019} P. S. Skardal and A. Arenas, {\em Phy. Rev. Lett.} {\bf 122}, 248301 (2019) 
	
	\bibitem{Xu2020} C. Xu, X. Wang and P. S. Skardal, {\em Phys. Rev. Res.} {\bf 2}, 023281 (2020)
	
	\bibitem{Skardal2020} P. S. Skardal and A. Arenas {\em Communication Physics} {\bf 3}, 218 (2020)
	\bibitem{Ajaydeep2022} A. D. Kachhvah and S. Jalan
{\em New J. of Phys.} (Fast Track) {\bf 24}, 052002 (2022).



\bibitem{Boccaletti2014} Zhang et.  al.,{\em Phys. Rev. Lett.}, {\bf 114} (3) (2014).

\bibitem{Pinaki2018}
P. Khanra, P. Kundu, C. Hens, and P. Pal,
{\em Phys. Rev. E}, {\bf  98}, 052315  (2018).

\bibitem{Anil_Inma2019}
S. Jalan, A. Kumar, I. Leyva
{\em Chaos: An Interdisciplinary Journal of Nonlinear Science} {\bf 29} (4), 041102
(2019)




\bibitem{Frolov2021}Frolov et. al., {\em Chaos, Solitons and Fractals},
{\bf 147}, 110955 (2021).

\bibitem{Anwar_Ghosh_2022} Md S. Anwar, D. Ghosh {\em Chaos: An Interdisciplinary Journal of Nonlinear Science} {\bf 32}, 033125 (2022).

\bibitem{Ghosh_chimera_2022}S. Kundu and D. Ghosh
{\em Phys. Rev. E} {\bf 105} L042202 (2022).

\bibitem{Kura1984} Y. Kuramoto, {\em International Symposium on Mathematical Problems in Theoretical Physics, Lecture Notes in Physics} {\bf 39} (1975).

\bibitem{Ott_Antonsen}
E. Ott and T. M. Antonsen, {\em Chaos : An Interdisciplinary Journal of Nonlinear Science} {\bf 18}, 037113 (2008).

\bibitem{Strogatz_Kura}S. H. Strogatz, {\em Physica D} {\bf 143}, 1 (2000).

\bibitem{supra_lap2013}A. S.-Ribalta {\it et. al.}, {\em Phys. Rev. E} {\bf 88} 032807 (2013).

	\bibitem{Ajaydeep_explosive_chimera2021}
A. D. Kachhvah, S. Jalan
{\em Phys. Rev. E} (Letter) 104 (4), L042301 (2021).
	\bibitem{Aoki2009} T. Aoki and T. Aoyagi, {\em Phys. Rev. Lett} {\bf 102}, 034101 (2009).
	
	\bibitem{Berner2020} R. Berner, J. Sawicki and E. Sch\"oll, {\em Phys. Rev. Lett.} {\bf 124} (8), 088301 (2020)
  
 	\bibitem{Berner2021} R. Berner, S. Vock, E. Sch\"oll and S. Yanchuk, {\em Phys. Rev. Lett.} {\bf 126}(2), 028301 (2021)

\bibitem{SJ_Ajaydeep_NJP2022}A. D. Kachhvah and S. Jalan, {\em New Journal of Physics} (Fast Track) {\bf 24}, 052002 (2022).


\bibitem{Karsai2021}
G. Cencetti, F. Battiston, B. Lepri and M. Karsai, {\em Sci. Rep.}  {\bf 11} 7028 (2021). 


\bibitem{commphy2021}    F. Musciotto, F. Battiston and R. N. Mantegna, {\em Comm. Phy.} {\bf 4}, 218 (2021).

\bibitem{rev_multi}A. Aleta and Y. Moreno, {\em Annual Review of Condensed Matter Physics}{\bf 10}, 45 (2019).

\bibitem{rev_higherorder}
F. Battistona {\it et. al.}, {\em {Physics Reports} {\bf }874}, 1 (2020).

\end{thebibliography}
\end{document}